\documentclass[conference]{IEEEtran}
\IEEEoverridecommandlockouts
\usepackage{cite}
\usepackage{amsmath,amssymb,amsfonts}
\usepackage{algorithmic}
\usepackage{graphicx}
\usepackage{textcomp}
\usepackage{xcolor}
\usepackage{enumitem}

\newcommand{\figureref}[1]{Fig.~\ref{#1}}
\def\BibTeX{{\rm B\kern-.05em{\sc i\kern-.025em b}\kern-.08em
    T\kern-.1667em\lower.7ex\hbox{E}\kern-.125emX}}
\begin{document}

\title{Decentralized Hybrid Precoding for Massive MU-MIMO ISAC\\
\thanks{This paper is supported in part by National Natural Science Foundation of China Program(62271316, 62101322), National Key R\&D Project of China (2019YFB1802703), Shanghai Key Laboratory of Digital Media Processing (STCSM 18DZ2270700) and the Fundamental Research Funds for the Central Universities.

The authors: Jun Zhu, Yin Xu, Dazhi He, Haoyang Li, YunFeng Guan and Wenjun Zhang are from Shanghai Jiao Tong University. Dazhi He is from Pengcheng Laboratory. The corresponding author is Dazhi He (e-mail: hedazhi@sjtu.edu.cn).
}
}

\author{
    \IEEEauthorblockN{Jun Zhu\IEEEauthorrefmark{1}, Yin Xu\IEEEauthorrefmark{1}, Dazhi He\IEEEauthorrefmark{1}\IEEEauthorrefmark{3}, Haoyang Li\IEEEauthorrefmark{1},  YunFeng Guan\IEEEauthorrefmark{2}, Wenjun Zhang\IEEEauthorrefmark{1}, \textit{Fellow, IEEE} }
    \IEEEauthorblockA{\IEEEauthorrefmark{1} Cooperative Medianet Innovation Center (CMIC), Shanghai Jiao Tong University\\\IEEEauthorrefmark{3}Pengcheng Laboratory \\Shanghai 200240, China \\ Email: \{zhujun\_22,  xuyin, hedazhi,  lihaoyang,  zhangwenjun\}@sjtu.edu.cn\\\IEEEauthorrefmark{2} Institute of Wireless Communication Technology, College of Electronic Information and Electrical Engineering \\ Email: yfguan69@sjtu.edu.cn}
}

\maketitle

 \begin{abstract}
Integrated sensing and communication (ISAC) is a very promising technology designed to provide both high rate communication capabilities and sensing capabilities. However, in Massive Multi User Multiple-Input Multiple-Output (Massive MU MIMO-ISAC) systems, the dense user access creates a serious multi-user interference (MUI) problem, leading to degradation of communication performance. To alleviate this problem, we propose a decentralized baseband processing (DBP) precoding method. We first model the MUI of dense user scenarios with minimizing Cramér-Rao bound (CRB) as an objective function. Hybrid precoding is an attractive ISAC technique, and hybrid precoding using Partially Connected Structures (PCS) can effectively reduce hardware cost and power consumption. We mitigate the MUI between dense users based on Thomlinson-Harashima Precoding (THP). We demonstrate the effectiveness of the proposed method through simulation experiments. Compared with the existing methods, it can effectively improve the communication data rates and energy efficiency in dense user access scenario, and reduce the hardware complexity of Massive MU MIMO-ISAC systems. The experimental results demonstrate the usefulness of our method for improving the MUI problem in ISAC systems for dense user access scenarios.
\end{abstract}
\section{Introduction}
\IEEEPARstart{I}{ntegrated} Sensing and Communication (ISAC) technology enables wireless communication and radar sensing to share hardware and spectrum resources. It is considered one of the most promising solutions to alleviate the growing spectrum shortages expected with the upcoming sixth-generation (6G) wireless systems\cite{b1}. Unlike conventional communication systems, ISAC effectively utilizes the same resources and hardware for both communication and sensing. In \cite{b2}, distributed multiple-input multiple-output (MIMO) access points are proposed to jointly serve and sense communication user. In \cite{b3}, this literature elaborates on the differences between the near-field and the original field of the base station and analyzes and represents the near-field ISAC. As the number of users increases, ISAC systems require more antenna units to further enhance communication capacity.

Massive Multi User Multiple-Input Multiple-Output (MU-MIMO) technology enables the simultaneous transmission of data streams from multiple users by leveraging multiple antennas among user devices. This technology significantly enhances spectral efficiency, capacity, and coverage of wireless communication systems, thereby providing users with more stable and high-speed communication services \cite{b4}, \cite{b5}. However, multi-user interference (MUI) in MIMO systems has been identified as an urgent problem in \cite{b6}.

In recent years, ISAC has attracted much attention in fields such as intelligent transportation and connected vehicles. These scenarios then transmit massive amounts of data while also requiring higher radar accuracy. In order to support the high communication data and radar targeting accuracy required for ISAC applications, millimetre wave can satisfy both functions. The major obstacle of high propagation path loss due to the high frequency of millimetre waves has to be overcome by using large scale antenna arrays to provide better gain. Large-scale transmitting antenna arrays can provide good gain, but are accompanied by complexity and multi-user interference problems that need to be addressed. In \cite{b7} partially connected hybrid precoding is used in ISAC which reduces the hardware complexity and mitigates the inter-user interference to some extent. In \cite{b8} and \cite {b9}, they use symbol level precoding to jointly design the transmission signal using channel knowledge and knowledge of the symbols to be transmitted. Although symbol precoding gives better gain in finite length data, it is not the best solution for multi-user interference (MUI) handling. In \cite {b10}, they adopt a method of using zero forcing (ZF) precoding under power constraints to reduce complexity and address multi-user interference. In addition to the analysis of single BS ISAC systems mentioned above, they studied the sensing performance of ISAC in the absence of cells in \cite {b11}. In summary, there has been some research on inter-user interference and complexity reduction, but none of the research is very mature.

Under the requirement of high rate, high performance communication, the model of ultra-large scale antenna size poses a significant challenge to the design of precoding schemes. The complexity of traditional linear precoding grows exponentially under ultra-large scale antennas, and the complexity of nonlinear precoding will become a catastrophic difficulty to be solved urgently. To address the challenges of precoding design bring about by ultra-large scale antennas, most schemes nowadays use Cell free geographically separated access points (APs) to jointly send coherent signals to all mobile users\cite{b12}. However, the cell free approach is not suitable for the situation of multiple targets and users in a single base station. Currently, research on single station multiple targets and users mostly focuses on low complexity linear precoding techniques in \cite{b10}. However, these precoding techniques are not the best solution to solve user interference and are mostly limited by the high complexity design of precoding. To address these issues, our research focuses on on a new baseband signal processing architecture, decentralized baseband processing (DBP) . Unlike the cell free scenario, it focuses on reducing the complexity of a single BS, which focuses on decentralizing center unit (CU) tasks to decentralized unit (DUs) for processing, thus reducing the complexity of CU.

In this paper, we focus on the perspective of counteracting the mulituser interference and reducing the system complexity, and we develop a MUI model for Massive MU MIMO-ISAC systems and establish an optimisation function with the objective of minimizing the Cramér-Rao bound (CRB). We reduce the MUI based on THP technique, adopt a DBP-based method to reduce the CU complexity and use partially linked architecture to reduce the hardware complexity. The  approach can effectively improve the communication performance of ISAC systems in idense user access scenarios. We demonstrate the feasibility and effectiveness of the proposed  approach through simulation experiments.

 The rest of the paper is composed as follows. In Section \text{II}, the system model is proposed to introduce the distributed partially connected hybrid precoding scheme, and the communication model and the sensing model are presented separately. In Section \text{III},  presents the precoding algorithm for star communication model, including centralized communication model and decentralized communication model, hybrid precoding, and the complexity of the algorithms is also analysed and these algorithms are compared.  In Section \text{IV}, simulation results are shown to verify the feasibility of our algorithm, and in Section \text{V}, the conclusions of this paper are summarized.
 
Notation: Bold uppercase letters denote matrices and bold lowercase letters denote vectors. For a matrix $\textbf A$, $\textbf A^T$, $\textbf A^H$, $\textbf A^{-1}$,  denote its transpose, conjugate transpose, inverse, respectively, blkdiag($\textbf H_1, ..., \textbf H_K$) denotes a block diagonal matrix with $\textbf H_1, ..., \textbf H_K$ being its diagonal blocks. The space of M × N complex matrices is expressed as $\mathbb{C}^{M\times N}$. Expectations are expressed as $\mathbb{E}[\textbf A]$.

\section{SYSTEM MODEL}

\begin{figure}[htbp]
\centering
\resizebox{0.45\textwidth}{!}{\includegraphics[]{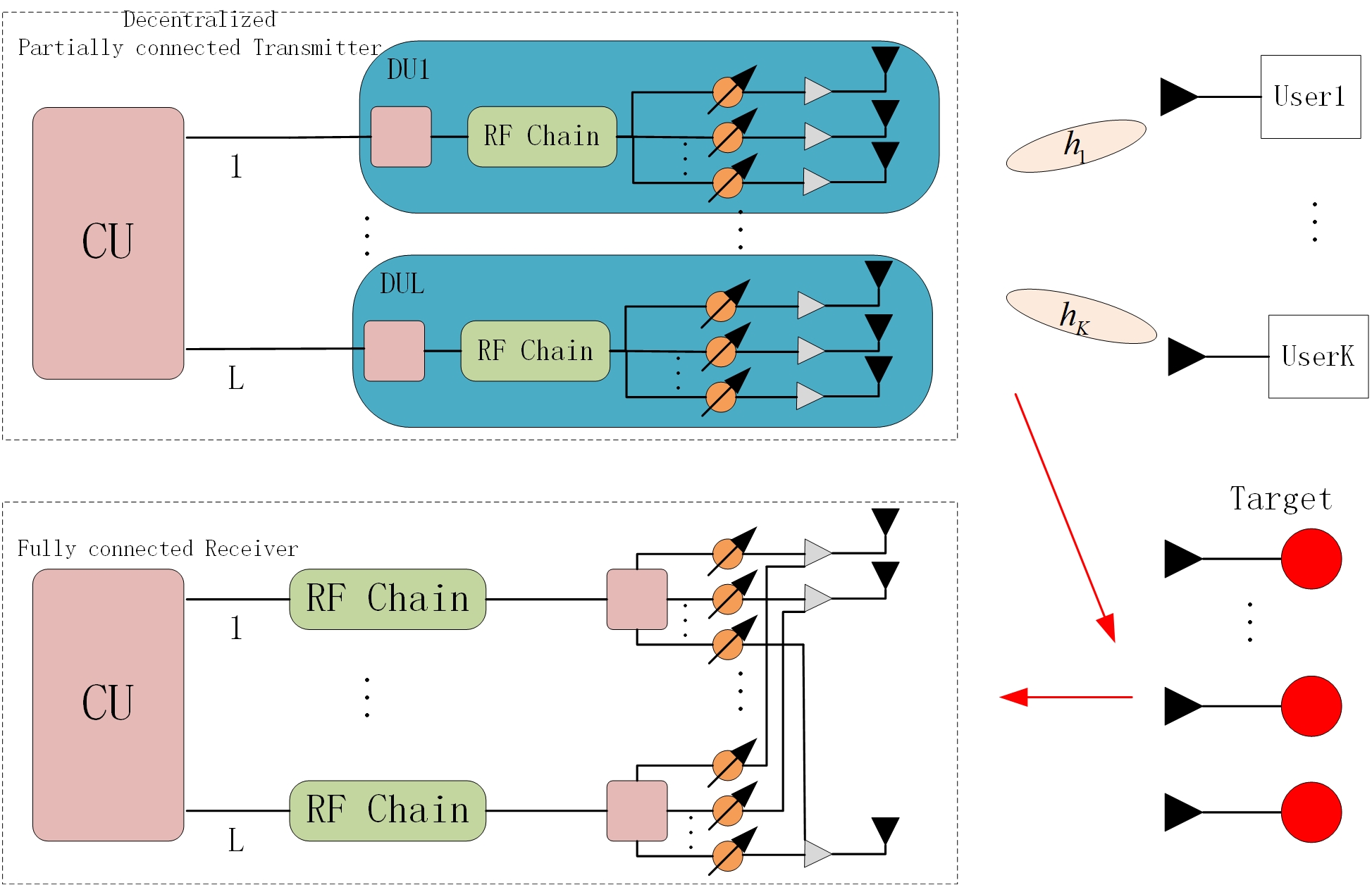}}
\caption{ISAC System.}
\label{fig:1}
\end{figure}

In this section, we provide an introduction to system models, including system model, communication model and sensing model. As shown in \figureref{fig:1}, we can see that there are a transmitter and a receiver at the BS. The transmitter uses partially connected hybrid precoding to reduce precoding design and hardware overhead of the RF chain. The receiver uses fully connected technology to distinguish the transmitter and cancel interference. The transmitter transmits information to the user and the target, and the target transmits it back to the BS receiver through perception.
\subsection{Communication Model}
As shown in \figureref{fig:1} the considered system for ISAC base station (BS) and $K$ single antenna users, in our system distributed hybrid precoding is used. We assume that the receiver is also equipped with an all-digital receiving  uniform linear array (ULA) in BS , spatially separated to a large extent from the transmitting array, in order to suppress interference directly from the transmitter. Note that this all-digital array can also be used to receive uplink communication signals, thereby improving uplink communication performance.
\begin{enumerate}[leftmargin=*] 
\setlength{\itemindent}{1em} 
    \item Centralized Communication Model
\end{enumerate}

Assume that the Base Station is equipped with $N_t$ transmit antennas and $N_r$ receive antennas in the downlink of a ISAC Massive MU-MIMO system. User $k$ is equipped with $N_k$ antennas. The total receiving antenna is $N = \sum_{k=1}^{K}N_k$.  $\textbf F_{RF,k}\in {\mathbb{C}^{N_t\times{N_k}}}$ and $\textbf F_{BB,k}\in {\mathbb{C}^{N_k\times{N_k}}}$ denote analog precoding matrix and digital precoding matrix respectively, transmits the signal $\textbf s_k\in {\mathbb{C}^{{N_k}\times1}}$, $\textbf x\in {\mathbb{C}^{{N_t}\times1}}$ indicates a precoding output signal, then  output signal $\textbf x$ is given by
\begin{equation}
\textbf x = \sum_{k=1}^{K} \textbf F_{RF,k} \textbf F_{BB,k}\textbf s_k\label{eq}.
\end{equation}
where $\textbf F_{RF} = [\textbf F_{RF,1}, \textbf F_{RF,2}, ..., \textbf F_{RF,K}]\in {\mathbb{C}^{N_t\times{N}}}$ and $\textbf F_{BB} = [\textbf F_{BB,1}, \textbf F_{BB,2}, ..., \textbf F_{BB,K}]\in {\mathbb{C}^{N\times{N}}}$.

In a flat fading channel, the received signal is denoted as
\begin{equation}
\begin{aligned}
    \textbf y_k &= \textbf H_k\textbf x+\textbf n_k \\
    & = \textbf H_k\textbf F_{RF,k}\textbf F_{BB,k}\textbf s_k+\sum_{j=1,j\neq k}^{K} \textbf H_k\textbf F_{RF,k}\textbf F_{BB,J}\textbf s_j+\textbf n_k,
\end{aligned}
\label{eq:2}
\end{equation}
where $\textbf y_k\in {\mathbb{C}^{{N_k}\times1}}$ is the receiving user vector received as user $k$, $\textbf H_k\in {\mathbb{C}^{{N_k}\times N_t}}$ denotes the channel matrix from the BS to the user k, $\textbf n_k\in {\mathbb{C}^{{N_k}\times1}}$ is the additive white Gaussian noise vector with distribution $\mathcal{CN}(0, \sigma_k^2 I)$, $\textbf H_k\textbf F_{RF,k}\textbf F_{BB,k} \textbf s_k$ denotes the information $\textbf s$ obtained by user k through channel, $\sum_{j=1,j\neq k}^{K} \textbf H_k\textbf F_{RF,j}\textbf F_{BB,j}\textbf s_j$ indicates that the signal received by user k is interfered with by other channels. $\textbf y \triangleq [\textbf y_1^T, \textbf y_2^T, ..., \textbf y_K^T]^T\in {\mathbb{C}^{{N}\times1}}$, $\textbf H\triangleq [\textbf H_1^T, \textbf H_2^T, ..., \textbf H_K^T]^T\in {\mathbb{C}^{{N}\times N_t}}$, $\textbf s\triangleq [\textbf s_1^T, \textbf s_2^T, ..., \textbf s_K^T]^T\in {\mathbb{C}^{{C}\times1}}$, $\textbf n\triangleq [\textbf n_1^T, \textbf n_2^T, ..., \textbf n_K^T]^T\in {\mathbb{C}^{{N}\times1}}$, Eq. \eqref {eq:2} can then be written in the following form
\begin{equation}
\begin{aligned}
     \textbf y &= \textbf H\textbf F_{RF}\textbf F_{BB}\textbf s+\textbf n ,
\end{aligned}
\label{eq:3}
\end{equation}
where the data streams are assumed to be independent of each other, so $\mathbb{E}[\textbf s \textbf s^H] = \textbf I$.

In our ISAC system, we use a distributed system model to reduce complexity. The advantage of the distributed approach is that nonlinear precoding can be used to achieve improved flux performance without increasing complexity.
\begin{enumerate}[leftmargin=*] 
\setlength{\itemindent}{1em} 
\setcounter{enumi}{1} 
    \item Dencentralized Communication Model
\end{enumerate}  

In the distributed ISAC system, we partition the user channel into $\textbf H_k = [\textbf H_k^1, \textbf H_k^2, ..., \textbf H_k^L]\in {\mathbb{C}^{{N_k}\times N_t}}$, the precoding matrix for user k is $\textbf F_{RF, k} = [ (\textbf F_{RF, k}^1)^T,  (\textbf F_{RF, k}^2)^T, ...,  (\textbf F_{RF, k}^L)^T]^T\in {\mathbb{C}^{N_t\times N_k }}$,  $\textbf F_{BB, k} = [ (\textbf F_{BB, k}^1)^T, (\textbf F_{BB, k}^2)^T, ...,  (\textbf F_{BB, k}^L)^T]^T\in {\mathbb{C}^{N_k\times{N_k}}}$. The BS transmit signal can be represented as $\textbf x = [ (\textbf x^1)^T,  (\textbf x^2)^T, ..., (\textbf x^l)^T]^T$. Consequently, the receive signal $y_k$ can be expressed as
\begin{equation}
\begin{aligned}
     \textbf y_k &= \sum_{l=1}^L\textbf H_k^l\textbf x^l+\textbf n_k \\
    & = \sum_{l=1}^L\textbf H_k^l\textbf F_{RF,k}^l\textbf F_{BB,k}^l\textbf s_k+\sum_{j\neq k}^{K}\sum_{l=1}^{L} \textbf H_k^l\textbf F_{RF,j}^l\textbf F_{BB,j}^l\textbf s_j+\textbf n_k,     
\end{aligned}
\label{eq:4}
\end{equation}
where $N_{l}$ is the number of antennas per unit that divides the $N_t$ antennas of the CU into $l$ DU units, $N_t = \sum_{l=1}^{L} N_{l}$ , $\textbf H_k^l\in {\mathbb{C}^{{N_k}\times N_{l}}}$ is the local channel,  $\textbf x^l\in{\mathbb{C}^{{N_l}\times 1}}$ is the local precoded signal to transmit, $\textbf F_{RF,k}^l\in{\mathbb{C}^{{N_{l}}\times N_k}}$ and $\textbf F_{BB,k}^l\in{\mathbb{C}^{{N_{l}}\times N_k}}$ denoted as analog precoding and digital precoding, respectively, for cluster $l$. When $l = 1$, DBP and CBP are equivalent. We can know that when the performance of decentralized and centralized is compared, the performance of centralized is generally higher than decentralized because the data information is more complete in centralized.

\begin{figure}[htbp]
\centering
\resizebox{0.5\textwidth}{!}{\includegraphics[]{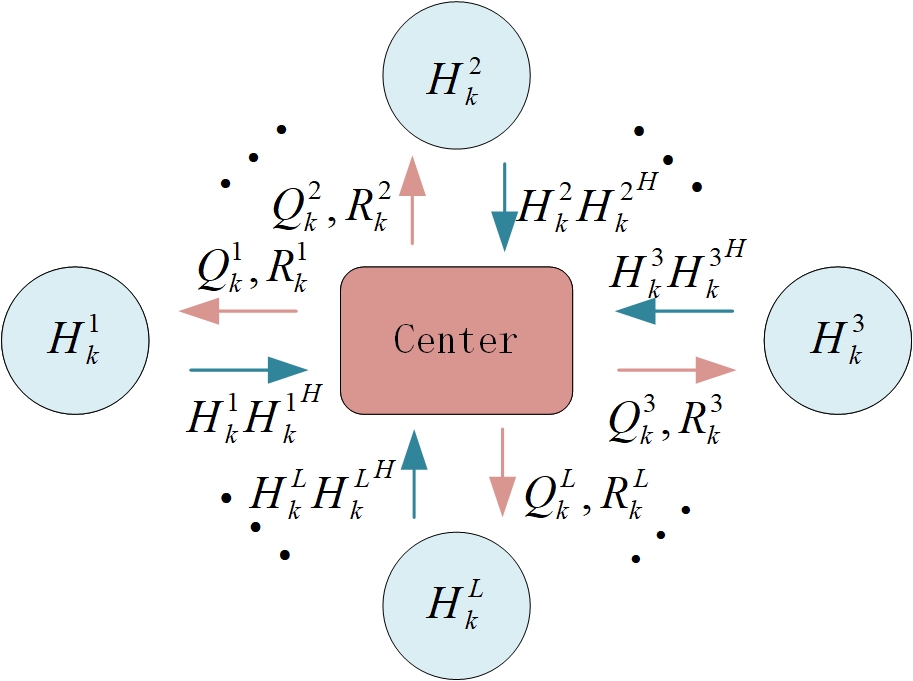}}
\caption{The communication and computation operations during the sDHMMSE-THP.}
\label{fig:2}
\end{figure}
According to the principle of star DBP, we place the linear computation in CU and the nonlinear computation in DUs, where CU receive channel state information (CSI) from DUs and transmit the relevant parameters back to DUs for nonlinear computation to finally determine the precoding\cite{b14}. The data transmission structure of the nonlinear precoding THP is shown in \figureref{fig:2}.
\subsection{Sensing Model}
In the perception signal model, the increase in inter user interference when serving high-density users can lead to a decrease in communication performance, indirectly causing a decrease in perception performance. Therefore, it is necessary to address inter user interference in order to provide better performance for perception.

Practically, there will be signal-related interference in MIMO perception. The sensing capabilities can be significantly enhanced by employing both analog and digital cancellation techniques, as outlined in \cite{b13}, effectively suppressing interference. Therefore, we assume that the noise term contains residual precoding, then the echo signal received at the BS can be represented as
\begin{equation}
\begin{aligned}
     \textbf Y &= \xi_0\textbf a_r(\psi_0)\textbf a_t^T(\psi_0)\textbf x+\sum_{t=1}^T \xi_t\textbf a_r(\psi_t)\textbf a_t^T(\psi_t)\textbf x+ \textbf N \\
    & = \xi_0\textbf A(\psi_0)\textbf x+\sum_{t=1}^T \xi_t\textbf A(\psi_t)\textbf x+ \textbf N ,
\end{aligned}
\label{eq:5}
\end{equation}
where $\xi_0$ and $\xi_t$ are the path loss and complex reflection coefficients corresponding to the target and the i-th interference, respectively, $\textbf a_r(\psi_0)\in{\mathbb{C}^{{N_r}\times 1}}$, $\textbf a_r(\psi_t)\in{\mathbb{C}^{{N_t}\times 1}}$, $\textbf N\in{\mathbb{C}^{{N_r}\times 1}}$ denote the direction vector of the transmitting antenna, the direction vector of the receiving antenna, and the covariance matrix with zero mean $\textbf R_N$ under AWGN, respectively. To evaluate the sensing performance, we use the CRB, which provides a lower bound for the mean square error of any unbiased (or asymptotically unbiased) parameter estimate. According to \cite{b13}, we can cancel out the clutter and obtain the CRB formula as

\begin{equation}
\begin{aligned}
      {CRB} &= \frac{1}{2|\xi_0|^2}(\textbf {Tr}(\textbf F_{BB}^H\textbf F_{RF}^H\dot{\textbf A}\textbf R_N^{-1}\dot{\textbf A}\textbf F_{RF}\textbf F_{BB}))^{-1}, 
\end{aligned}
\label{eq:6}
\end{equation}
where $\dot {\textbf A} = \dot {\textbf a}(\psi) {\textbf a}^T(\psi)+ {\textbf a}(\psi)\dot {\textbf a}^T(\psi)$, $\dot {\textbf a}(\psi)$ denotes the derivative of the vector.

\begin{equation}
\begin{aligned}
      \dot {\textbf a}(\psi) &= [0, j2\pi\delta a_2cos(\psi), ..., j2\pi\delta(N-1) a_Ncos(\psi)], 
\end{aligned}
\label{eq:7}
\end{equation}
where $a_i$ denotes the i-th location of the $\textbf a(\psi)$.
\subsection{Problem Formulation}
In this subsection, we present the problem with the goal of co-designing decentralized analog precoding $\textbf F_{RF}$ and digital precoding $\textbf F_{BB}$ and  to minimise the CRB, and the equation can be expressed as

\begin{equation}
\begin{aligned}
     \underset{\textbf  F_{BB},\textbf F_{RF}}{\min}  &CRB, \\
      &\text{s.t.} \quad ||\textbf F_{RF}\textbf F_{BB}|| = P,      
\end{aligned}
\label{eq:8}
\end{equation}
which minimizes the CRB while limiting the power, the formula for the CRB can be obtained from \eqref{eq:6}. By minimizing the CRB with respect to a particular direction, we can improve sensing in that desired sensing direction. Based on (eqref{eq:6}), we know that $\textbf F_{RF}\textbf F_{BB}$ aims to minimize MUI interference. Therefore, Eq \eqref{eq:8} can be rewritten as

\begin{equation}
\begin{aligned}
     \underset{\textbf  F_{BB},\textbf F_{RF}}{\min}  \underset{}{\max} &\frac{1}{2|\xi_0|^2}(\textbf {Tr}(\textbf F_{BB}^H\textbf F_{RF}^H\dot{\textbf A}\textbf R_N^{-1}\dot{\textbf A}\textbf F_{RF}\textbf F_{BB}))^{-1}, \\
      &\text{s.t.} \quad ||\textbf F_{RF}\textbf F_{BB}|| = P,      
\end{aligned}
\label{eq:9}
\end{equation}
\section{PROPOSED ALGORITHM}
In this section, we will focus on introducing the basic principles of applying star decentralized mixed zero forcing Thomlinson Harashima precoding (sDHZF THP) and star de targeting mixed minimum mean square error Thomlinson Harashima precoding (sDHMMSE THP), as well as the design of mixed precoding. The purpose of these two algorithms is to reduce user interference and lower CRB.
\subsection{Star Decentralized  Precoding}
In ISAC MUI system, the handling of MUI is very important and THP has better results in handling MUI. THP usually uses three filters, namely feedforward filter $\textbf F\in {\mathbb{C}^{{N_t}\times N}}$, which partially removes MUI, feedback filter $\textbf B\in {\mathbb{C}^{{N}\times N}}$, which is a lower triangular matrix, and weighting matrix $\textbf G\in {\mathbb{C}^{{N}\times N}}$, which contains weighting factors for each stream.

Based on \figureref{fig:2}, we know that each DU transmits $\textbf H_k^l$ to CU, the QR decomposition of  $\textbf H_k$ shown below
\begin{equation}
\begin{aligned}
     {{\textbf H_k}^H} &= \textbf Q_k\textbf R_k,  \forall k,
\end{aligned}
\label{eq:9}
\end{equation}
where  $\textbf Q_k\in {\mathbb{C}^{{N_t}\times N_k}}$ is the unitary matrix and $\textbf R_k\in {\mathbb{C}^{{N_k}\times N_k}}$ is the upper triangular matrix, along with the weighting matrix $\textbf G$. In the classical ZF-THP, the feedforward matrix $\textbf F_k^l\in {\mathbb{C}^{{N_l}\times N_k}}$  is set as $\textbf Q_k^l$. Each DU receives $\textbf R_k^l\in {\mathbb{C}^{{N_k}\times N_k}}$, $\textbf Q_k^l$ and calculates $({\textbf B_k^l})^{-1} = ({{\textbf R_k^l}^H})^{-1}{\textbf G_k^l}^{-1} $. $\textbf G_k^l$ is as follows
\begin{equation}
\begin{aligned}
   \textbf G_k^l = 
\begin{bmatrix}
r_{1,1}^{-1}   \\
&r_{2,2}^{-1}  \\
&& \ddots \\
&&&r_{N_k,N_{k}}^{-1} \\
\end{bmatrix}.
\end{aligned}
\label{eq:10}
\end{equation}

 The sDHZF-THP, we can rewrite Equation. \eqref {eq:5} as follows
\begin{equation}
\begin{aligned}
    \textbf y_k &=\textbf H_k\textbf x+\textbf n_k \\
    & = \sum_{l=1}^L\textbf G_k^l\textbf H_k^l\textbf F_{RF,k}^l\textbf F_{BB,k}^l({\textbf B_k^l})^{-1}\textbf s_k^l\\
    &+\sum_{j\neq k}^{K}\sum_{l=1}^{L} \textbf G_j^l\textbf H_k^l\textbf F_{RF,j}^l\textbf F_{BB,j}^l({\textbf  B_k^l})^{-1}\textbf 
 s_j^l+\sum_{l=1}^{L}\textbf G_k^l\textbf n_k,     \forall k.
\end{aligned}
\label{eq:11}
\end{equation}

Through the F, G and B filters in the THP, Equation. \eqref{eq:12} can rewrite  as:
\begin{equation}
\begin{aligned}
     \textbf y_k &= \textbf s_k+\sum_{l=1}^{L}\textbf G_k^l\textbf n_k. \\
\end{aligned}
\label{eq:12}
\end{equation}

According to Eq. \eqref{eq:12}, it is evident that sDHZF-THP retains a certain level of interference. To further minimize this interference, reduce the CRB and enhance communication transmission rates, the subsequent subsection introduces the sDHMMSE-THP method.

We utilize $\textbf H\textbf H^H$ as the channel transmission matrix between CU and DUs, which is independent of the total number of antenna transmitting antennas and greatly improves the system of performance. This not only reduces the computational overhead of CUs, but also reduces the computational overhead of CUs and DUs.

The communication and computation operations during the sDHMMSE-THP as shown in  \figureref{fig:2}, which $\textbf H_k^l{\textbf H_k^l}^H$ is transferred from the DU to the CU, which reduces the complexity considerably. In CU, each cluster $l$ to obtain $\textbf H_k \textbf H_k^H =\sum_{l=1}^L\textbf H_k^l{\textbf H_k^l}^H$, we know the following equation:
\begin{equation}
\begin{aligned}
     {\textbf H_k}^{-1}({\textbf H_k}{\textbf H_k}^H+ \xi\textbf I)^H &= \textbf Q_k\textbf R_k ,\\
\end{aligned}
\label{eq:13}
\end{equation}
where $\xi=\sigma_n^2 / \sigma_x^2$ as shown in \figureref{fig:1}, according to Equation. \eqref{eq:11}, we get the following formula: $\textbf F_k^l =\textbf Q_k^l$, each DU receives  $\textbf R_k^l$, $\textbf Q_k^l$  and calculates $({\textbf B_k^l})^{-1}$, $\textbf G_k^l=diag(r_{1,1}^{-1}, r_{2,2}^{-1}, ..., r_{N_k,N_k}^{-1})$. The sDMMSE-THP output signal is shown in Equation \eqref{eq:11}.
\subsection{Hybrid Precoding}
 The benefit of hybrid precoding is to reduce the hardware overhead but in case of distributed precoding it will increase the hardware overhead then selecting partial connectivity mode of hybrid precoding is feasible in our system.
Hybrid precoding is applied to cope with the overhead growth of all-digital hardware cost due to millimeter-wave transmission, and partial connectivity reduces the hardware cost overhead in the distributed case we propose. In our system analog precoding can only be phase optimised, if the phases are of finite b-bit resolution, so $\textbf F_{RF}$ can be expressed as $\textbf F_{RF} = \{e^{j\theta}|\theta \in{\{0, \frac{2\pi}{2^b}, ..., \frac{(2b-1)2\pi}{2^b}\}}\}$.

We can observe that the rate serves as a reference metric for performance,  which can be expressed as $R_k$, representing the rate at which the user k transmits a message, as shown in the following equation
\begin{equation}
\begin{aligned}
     &R_k \triangleq log\ det(\textbf I+\textbf H_k\textbf F_{RF,k}\textbf F_{BB,k}(\textbf H_k\textbf F_{RF,k}\textbf F_{BB,k})^H) \\
         &\quad \times \left(\sum_{j \neq k}(\sigma_k^2 \textbf I+\textbf H_k\textbf F_{RF,j}\textbf F_{BB,j}(\textbf H_k\textbf F_{RF,j}\textbf F_{BB,j})^H)\right)^{-1},
\end{aligned}
\label{eq:14}
\end{equation}
where using the formula of Shannon, $\textbf H_k\textbf F_{RF,k}\textbf F_{BB,k}(\textbf H_k\textbf F_{RF,k}\textbf F_{BB,k})^H$ denotes the useful information of k users and $\sum_{j \neq k}(\sigma_k^2 \textbf I+\textbf H_k\textbf F_{RF,j}\textbf F_{BB,j}(\textbf H_k \textbf F_{RF,j}\textbf F_{BB,j})^H$ denotes  noise and the noise interference with other channels.
\subsection{Complexity Analysis}
In this subsection, we highlight the low complexity of our distributed approach by comparing it with the low-complexity ZF precoding method proposed in \cite{b10}. This comparison illustrates the superiority of distributed precoding in terms of complexity. We analyze and compare the complexities of CZF, CHZF-THP, sDHZF-THP, and sDHMMSE-THP using floating point operations (FLOPS) as a metric for computation:

\textbullet \ $\textbf H\textbf H^H, \textbf H\in {\mathbb{C}^{{n}\times m}}$ requires $nm^2$ FLOPS.      

\textbullet \ The QR decomposition of $\textbf H$ requires $\frac{2}{3}n^3+mn^2$ FLOPS.

 Combining Equation \eqref{eq:2}, classic CZF needs $2nm^2$ FLOPS and Classic CHZF-THP uses QR decomposition to obtain $\frac{2}{3}n^3+m^2+2mn^2+mn+m$ FLOPS,according to Equation \eqref{eq:5}, by using sDHZF-THP in the CU requires $\frac{2}{3}n^3+mn^2+mn$ FLOPS. Combining \eqref{eq:5}, \eqref{eq:10} and the classical MMSE THP algorithm, by using sDHMMSE-THP in the CU requires $\frac{2}{3}n^3+mn^2+mn+m$ FLOPS. 
\begin{figure}[htbp]
\centerline{\includegraphics[width=0.5\textwidth]
{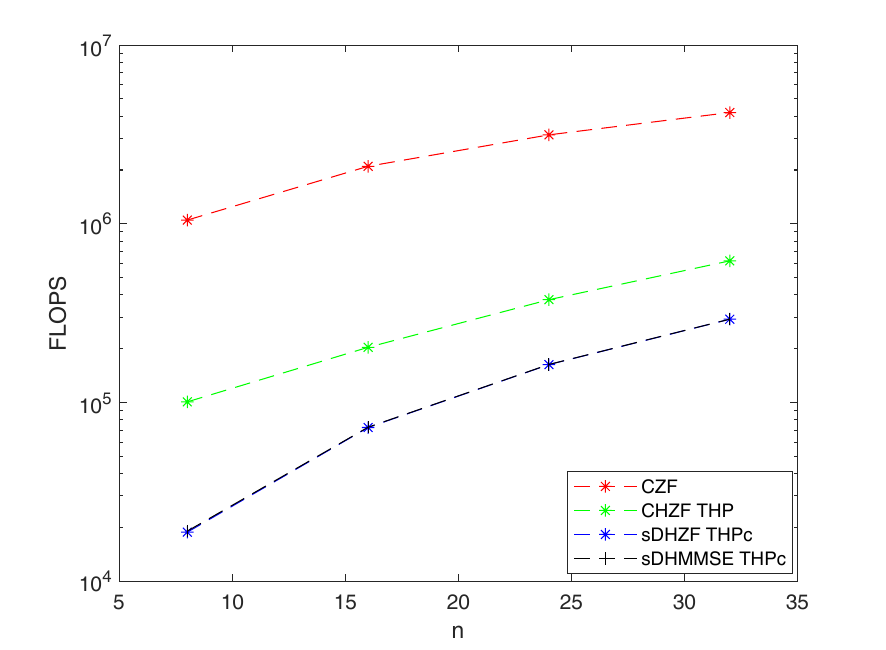}}
\caption{Complexity analysis in terms of FLOPS with $N_t$=256.}
\label{fig:3}
\end{figure}

According to \figureref{fig:3}, we can see that sDHZF-THPc reduces the complexity by 52.8\% and 92.8\% compared to CZF and CHZF-THP, and the complexity of sDHMMSE-THP is slightly higher than sDHZF-THP, so our proposed sDHZF-THP and sDHMMSE-THP is beneficial to reduce the center complexity. 

The proposed algorithm reduces the CU complexity for different system sizes, all the required FLOPS using different algorithms are shown in  \figureref{fig:3}. sDMMSE-THP and sDZF-THP reduce the intermediate complexity compared to centralized, and sDHMMSE-THP improves the system performance while sacrificing some complexity compared to sDHZF-THP.
\section{SIMULATION RESULTS}
In this paper, we focus on the application of our proposed distributed algorithm in ISAC system, comparing the optimized CRB performance with hybrid precoding and comparing the total rate and communication performance. The simulation results show that we have achieved better results in improving the perceptual performance while optimizing the communication performance. The simulation comparison shows that we have achieved the total rate while improving the perceptual performance, indicating that the perceptual performance is good while improving the communication performance.

\begin{figure}[htbp]
\centerline{\includegraphics[width=0.5\textwidth]
{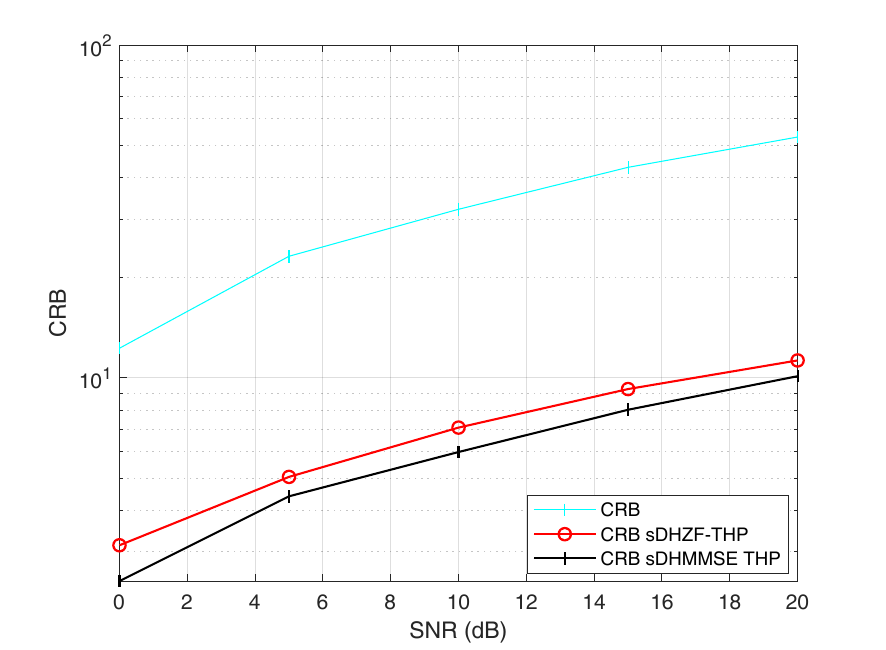}}
\caption{CRB of a Massive MU-MIMO ISAC system under Rayleigh channels($N_t=256, K=32, N_k=1, L=4$).}
\label{fig:4}
\end{figure}
In the first example, as shown in \figureref{fig:4},  under Rayleigh channel, we can see that our proposed algorithm has lower CRB and better performance compared to \cite{b10}.   We consider that the BS is configured with $N_t = 256$ transmitting antennas to broadcast data to $K = 32$ users, each user is configured with $N_k = 1$ receiving antenna, and the DUs are divided into $L = 4$ clusters. Simulation shows that our proposed sDHZF-THP and sDHMMSE-THP have lower and better CRB performance compared to the ZF mentioned in \cite{b10}.

\begin{figure}[htbp]
\centerline{\includegraphics[width=0.5\textwidth]
{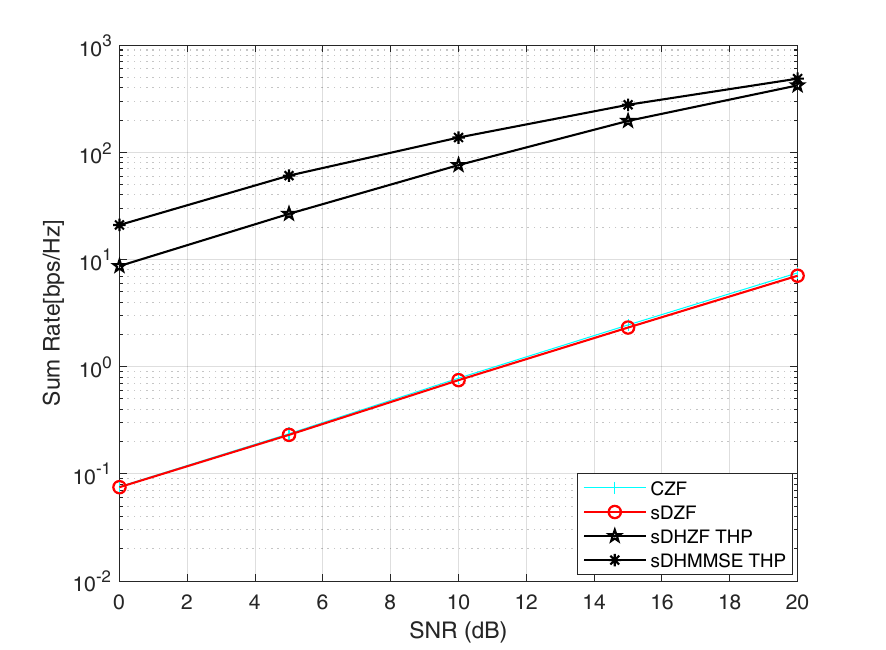}}
\caption{Sum-rate performance of a Massive MU-MIMO ISAC system employing different
 precoding schemes($N_t=256, K=32, N_k=1, L=4$).}
\label{fig:5}
\end{figure}

In the second example, shown in \figureref{fig:5}, we evaluate the total rate performance of the proposed algorithm. We consider that the BS is configured with $N_t = 256$ transmitting antennas to broadcast data to $K = 32$ users, each user is configured with $N_k = 1$ receiving antenna, and the DUs are divided into $L = 4$ clusters. Compared with CZF in \cite{b10}, our proposed sDHZF-THP and sDHMMSE-THP have higher total transmission rates. According to \figureref{fig:3}, sDHMMSE-THP further improves the system performance with unavoidable loss of complexity, and these results confirm that our newly proposed decentralized precoding scheme outperforms existing schemes

Unsurprisingly, sDHZF-THP and sDHMMSE-THP outperform the traditional decentralized CZF in terms of CRB performance and summed data transfer speed, in addition to demonstrating the effectiveness of the star DBP in reducing CU complexity. The advantage of star DBP over traditional decentralized CZF is that it can reduce the complexity of CU. In order to further improve the system performance, a certain amount of complexity needs to be lost, and sDHMMSE-THP can achieve this.

\section{CONCLUSION}
In this paper, we investigate hybrid precoding for ISAC scenarios, where the need to transmit large volumes of data necessitates the adoption of distributed precoding techniques. These techniques help reduce the complexity overhead associated with utilizing very large antenna arrays and managing multi-user interference. By employing a distributed system, hardware overhead can be minimized through partially connected hybrid precoding. Our proposed methods, sDHZF-THP and sDHMMSE-THP, effectively address this challenge, achieving low distributed complexity while maintaining high communication performance.


\vspace{12pt}
\color{red}

\end{document}